\documentclass[prd,aps,twocolumn,preprintnumbers, showpacs, nofootinbib,superscriptaddress,notitlepage]{revtex4-1}
\usepackage{amssymb, amsthm}
\usepackage{amsmath}    % subequations
\usepackage{mathrsfs}   % \mathscr
\usepackage{graphicx}   % figures
\usepackage{color}      % color is used in text
\usepackage{footnote}   % footnote in tabular environment, must be loaded after color
\usepackage{slashed}    % Feynman slash
\usepackage{hyperref}   % hypertext links, including those to external documents and URLs

\usepackage[utf8]{inputenc}
\usepackage[dvipsnames]{xcolor}
\usepackage[normalem]{ulem}

\newcommand{\nn}{\nonumber}

\begin{document}

\preprint{MSUHEP-18-019,MIT-CTP/5033}

\title{Nucleon Transversity Distribution at the Physical Pion Mass from Lattice QCD}

\collaboration{\bf{Lattice Parton Physics Project ($\rm {\bf LP^3}$) Collaboration}} 

\author{Yu-Sheng Liu}
\affiliation{Tsung-Dao Lee Institute, Shanghai Jiao-Tong University, Shanghai 200240, China}

\author{Jiunn-Wei Chen}
\affiliation{Department of Physics, Center for Theoretical Physics, and Leung Center for Cosmology and Particle Astrophysics, National Taiwan University, Taipei, Taiwan 106}
\affiliation{Center for Theoretical Physics, Massachusetts Institute of Technology, Cambridge, MA 02139, USA}

\author{Luchang Jin}
\affiliation{Physics Department, University of Connecticut,
Storrs, Connecticut 06269-3046, USA}
\affiliation{RIKEN BNL Research Center, Brookhaven National Laboratory,
Upton, NY 11973, USA}

\author{Ruizi Li}
\affiliation{Department of Physics and Astronomy, Michigan State University, East Lansing, MI 48824}

\author{Huey-Wen Lin}
\email{hwlin@pa.msu.edu}
\affiliation{Department of Physics and Astronomy, Michigan State University, East Lansing, MI 48824}
\affiliation{Department of Computational Mathematics, Science \& Engineering, Michigan State University, East Lansing, MI 48824}

\author{Yi-Bo Yang}
\affiliation{Department of Physics and Astronomy, Michigan State University, East Lansing, MI 48824}
\affiliation{Institute of Theoretical Physics, Chinese Academy of Sciences, Beijing 100190, China}

\author{Jian-Hui Zhang}
\affiliation{Institut f\"ur Theoretische Physik, Universit\"at Regensburg, D-93040 Regensburg, Germany}

\author{Yong Zhao}
\email{yzhaoqcd@mit.edu}
\affiliation{Center for Theoretical Physics, Massachusetts Institute of Technology, Cambridge, MA 02139, USA}

%\date{}

\begin{abstract}

We report a state-of-the-art lattice calculation of the isovector quark transversity distribution of the proton at the physical pion mass. Within the framework of large-momentum effective theory (LaMET), we compute the transversity quasi-distributions using clover valence fermions on 2+1+1-flavor (up/down, strange, charm) HISQ-lattice configurations with boosted proton momenta as large as 3.0~GeV. 
The relevant lattice matrix elements are nonperturbatively renormalized in regularization-independent momentum-subtraction (RI/MOM) scheme and systematically matched to the physical transversity distribution. 
With high statistics, large proton momenta and meticulous control of excited-state contamination, 
we provide the best theoretical prediction for the large-$x$ isovector quark transversity distribution, with better precision than  
the most recent global analyses of experimental data. 
Our result also shows that the sea quark asymmetry in the proton transversity distribution is consistent with zero, which has been assumed in all current global analyses. 

\end{abstract}
\maketitle

%%%%%%%%%%%%%%%%%%%%%%%%%%%%%%%%%%%%%%%%%%%%%%%%%%%%%%%%%%%%%%
{\em Introduction:}
The spin structure of the nucleon has been an important subject in particle and nuclear physics since first experiments on deep inelastic scattering (DIS) in the '60s.
Significant experimental efforts have greatly improved our knowledge of the spin structure over the last half century. 
However, the transversely polarized structure of the nucleon remains a puzzle. 
The simplest transverse structure is the nucleon transversity parton distribution function (PDF) $\delta q(x)$ and it has long been a focal point to study in hadronic community. 
As a chiral-odd quantity~\cite{Jaffe:1991ra}, $\delta q(x)$ can be accessed through the transverse-transverse spin asymmetry in Drell-Yan processes~\cite{Ralston:1979ys} 
or in the Collins single-spin asymmetry in semi-inclusive deep inelastic scattering (SIDIS), where it couples to the chiral-odd Collins fragmentation function~\cite{Collins:1992kk}.  
There are attempts in global analysis to constrain the transversity PDF from experimental data~\cite{Anselmino:2007fs,Anselmino:2008jk,Anselmino:2013vqa,Kang:2014zza,Kang:2015msa,Bacchetta:2011ip,Bacchetta:2012ty,Radici:2015mwa,Lin:2017stx,Radici:2018iag}.
However, the theoretical foundation for extracting transversity PDF from global experimental data is not yet fully established. Due to the small kinematic coverage of current experiments, the resulting distribution at small $x$ and $x>0.5$ is mainly extrapolation and heavily relies on the ansatz for the PDF. Therefore, $\delta q(x)$ from current global analyses still has a very large uncertainty.
Ongoing and future experiments, such as Jefferson Lab 12-GeV and SoLID, are expected to shed more light on the large-$x$ region and will greatly improve the transversity PDF determination.

In addition to studying the quark distribution, there has also been great interest in studying the flavor asymmetry in the nucleon sea-quark distributions (see, e.g.,~\cite{Chang:2014jba} for a review). The experimental observation of flavor asymmetry in the unpolarized and longitudinally polarized sea provides strong constraints on a wide range of QCD models. In terms of the transversely polarized nucleon sea, there has been some speculation that its size is likely to be smaller than the longitudinally polarized sea, and all current global analyses have taken every antiquark distribution to be exactly zero everywhere. 
Unfortunately, there is no experimental data to determine the size of such quantities, nor even the sign of the transverse sea-flavor asymmetry. 
Future experiments, such as the Electron-Ion Collider (EIC) or the Drell-Yan experiment at FNAL (E1027+E1039), may be able to unravel the mystery of the transversely polarized sea. 

The lack of a precise measurement of the transversity PDF highlights the necessity and usefulness of theoretical prediction
from first-principles methods such as lattice QCD. 
For decades, only the lowest moments of PDFs were calculated, so the transversity PDF could not be reconstructed from lattice studies directly. 
The status quo has been fundamentally changed by large-momentum effective theory (LaMET)~\cite{Ji:2013dva,Ji:2014gla}, a powerful approach that was proposed not long ago 
to calculate the Bjorken-$x$ dependence of many parton physics observables directly
from lattice QCD. In LaMET, one starts from the lattice matrix element of a Euclidean quasi-observable (``quasi-PDF'' for the calculation of PDFs) in a large-momentum hadron state, and then match it onto the corresponding parton observable through a factorization formula. In recent years, much progress has been made in both lattice and perturbative QCD on the theoretical development of LaMET~\cite{Ji:2017oey,Ishikawa:2017faj,Green:2017xeu,Alexandrou:2017huk,Chen:2017mzz,Zhang:2018diq,Li:2018tpe,Xiong:2013bka,Constantinou:2017sej,Stewart:2017tvs,Izubuchi:2018srq,Liu:2018uuj}. 
Meanwhile, LaMET has been applied to calculating the isovector quark PDFs, meson distribution amplitudes~\cite{Lin:2014zya,Alexandrou:2015rja,Chen:2016utp,Alexandrou:2016jqi,Zhang:2017bzy,Chen:2017gck,Alexandrou:2018pbm,Alexandrou:2018eet,Chen:2018xof,Chen:2018fwa,Lin:2018qky} and the total gluon polarization~\cite{Yang:2016plb}. 
The pioneering works on transversity first done by LP$^3$~\cite{Chen:2016utp} and then by ETMC~\cite{Alexandrou:2016jqi} were carried out at heavier pion masses without lattice renormalization. Very recently, ETMC reported a high-statistics calculation at the physical point with nucleon momentum of 1.4~GeV and the complete renormalization and matching procedures~\cite{Alexandrou:2018eet}.
However, recent studies by LP$^3$~\cite{Chen:2018fwa,Lin:2018qky} show that larger nucleon boost momenta are necessary to obtain the correct small-$x$ distribution and show the sign of the sea-flavor asymmetry.

In this work, we present a state-of-the-art lattice calculation of the isovector quark transversity PDF $\delta u(x)-\delta d(x)$ of the proton. The lattice matrix element of the transversity quasi-PDF is calculated at the physical pion mass with large nucleon momenta up to 3.0~GeV, renormalized in the regularization-independent momentum-subtraction (RI/MOM) scheme as elaborated in Refs.~\cite{Stewart:2017tvs,Chen:2017mzz,Liu:2018uuj}, and eventually matched onto the $\overline{\text{MS}}$ transversity PDF with newly derived one-loop matching coefficient in the Appendix. To remove excited-state contamination, we perform multi-state analyses using high-statistics data at six source-sink separations. The final result has reached a precision that is significantly better than the recent global analysis by the JAM collaboration (JAM17)~\cite{Lin:2017stx}, and is consistent with another fit (LMPSS17) constrained by the lattice-averaged tensor charge $g_T(\equiv \int dx\delta q(x))$~\cite{Lin:2017stx}. Our prediction of the transversity distribution at large $x$ and the small sea-quark flavor asymmetry (consistent with zero within errors) will have a significant impact on phenomenological studies.

%%%%%%%%%%%%%%%%%%%%%%%%%%%%%%%%%%%%%%%%%%%%%%%%%%%%%%%%%%
{\em Numerical calculation:}
The calculation is carried out with clover valence fermions on an ensemble with lattice spacing $a=0.09$~fm, box
size $L\approx 5.8$~fm, pion mass $M_\pi \approx 135$~MeV, and $N_f=2+1+1$ (degenerate up/down, strange and charm) dynamical flavors of highly improved staggered quarks (HISQ)~\cite{Follana:2006rc} generated by MILC Collaboration~\cite{Bazavov:2012xda}. We use Gaussian momentum smearing~\cite{Bali:2016lva} for the quark field to reach large proton boost momenta with $\vec{P}=\{0,0, n \frac{2\pi}{L}\}$ and $n \in \{10,12,14\}$, corresponding to 2.2, 2.6 and 3.0~GeV, respectively.  More details of the lattice setup and parameters can be found in Ref.~\cite{Lin:2018qky}.

\begin{figure}[thbp]
\includegraphics[width=.4\textwidth]{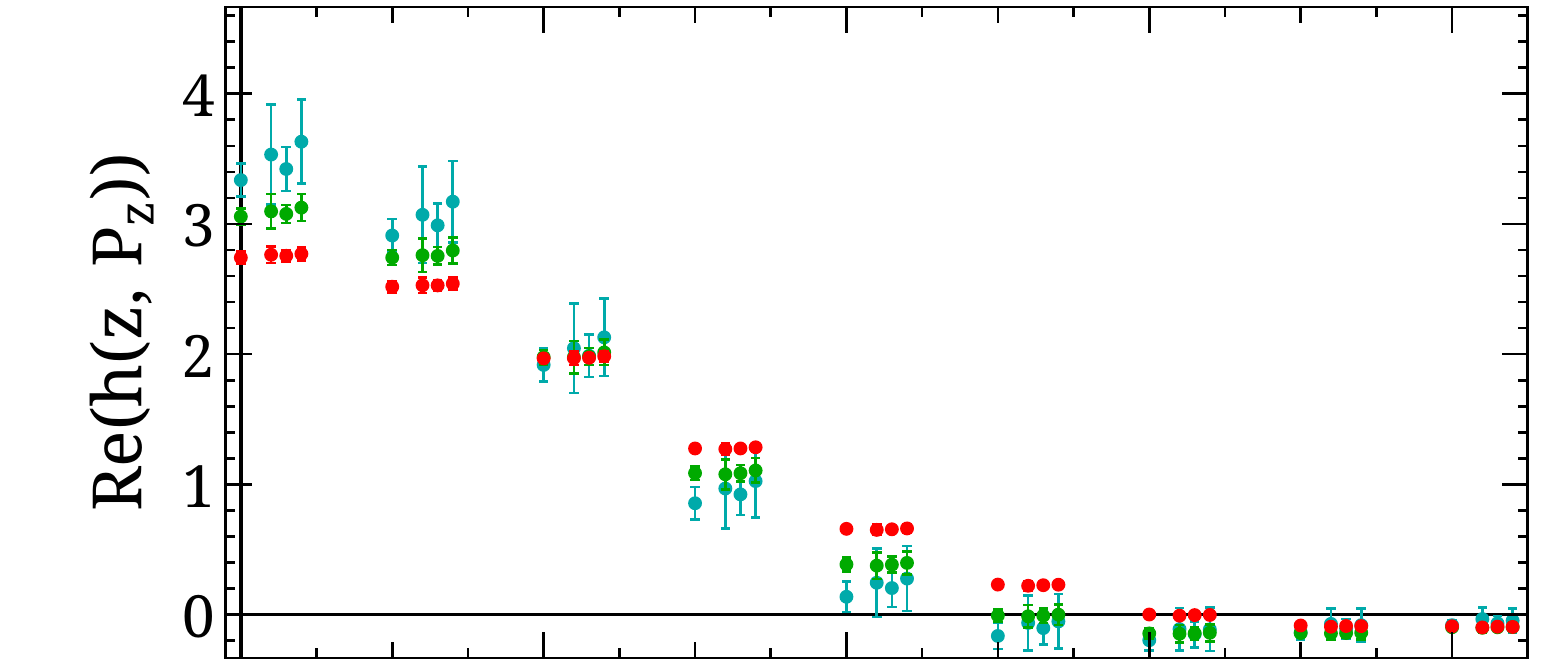}
\includegraphics[width=.4\textwidth]{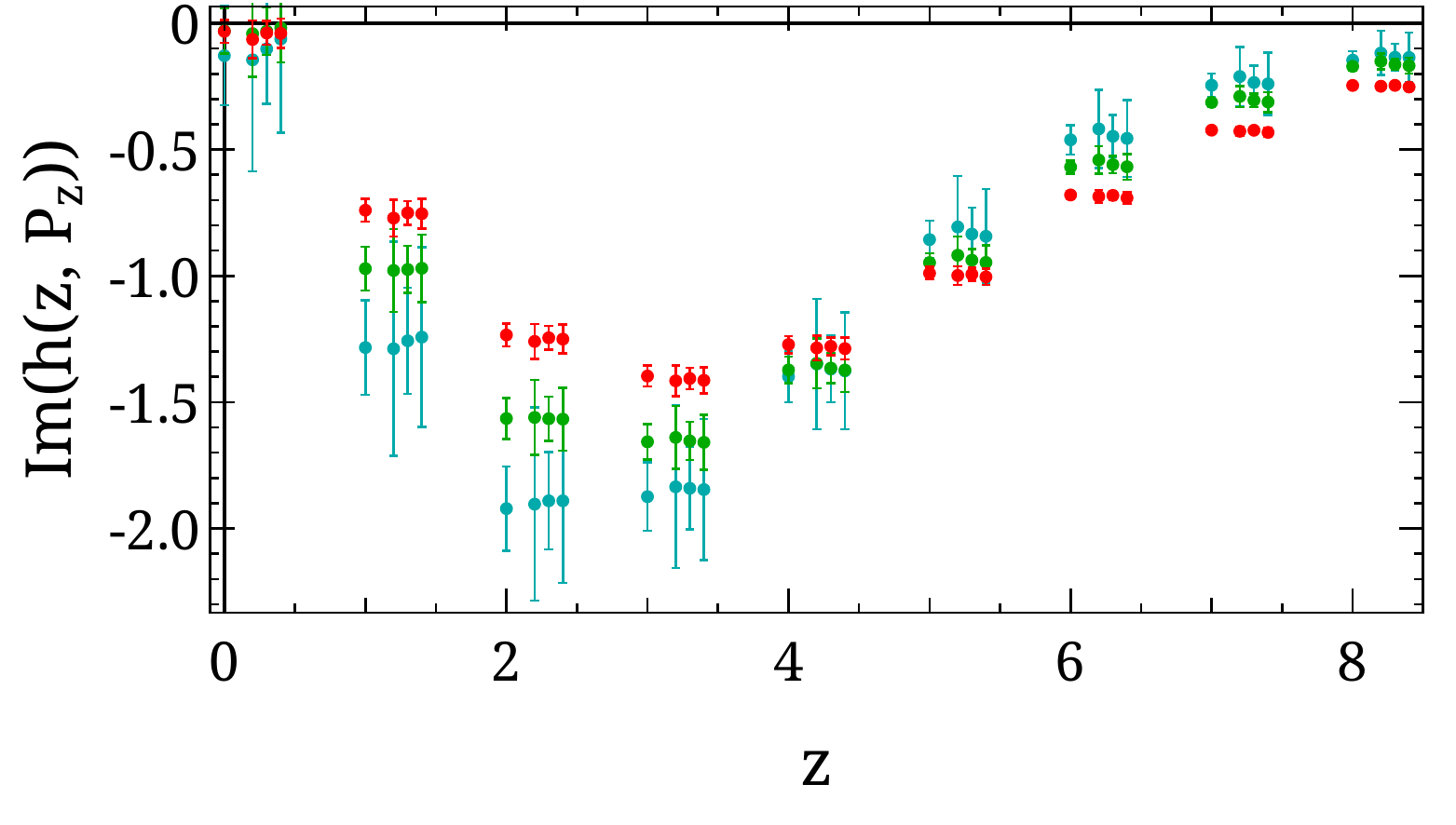}
\caption{
The scaled real (top) and imaginary (bottom) parts of the bare nucleon matrix elements for the isovector transversity quasi-PDFs as functions of $z$ at all three momenta (2.2, and 2.6 and 3.0~GeV indicated by red, green and blue, respectively).
We multiply different proton momenta by different factors to enhance visibility. At a given positive $z$ value, the data are slightly offset to show different ground-state extraction strategies; from left to right they are: two-simRR using all $t_\text{sep}$, two-simRR using the largest 4 $t_\text{sep}$, two-sim using the largest 3 $t_\text{sep}$, and two-sim using the largest 2 $t_\text{sep}$. Different analyses are consistent within statistical errors, which suggests the excited-state contamination is well controlled. 
} 
\label{fig:bareME-tsep}
\end{figure}

On the lattice, we first calculate
 the equal-time three-point correlator along the $z$-axis with operator $\hat{O}(z,a)=\bar{\psi}_q(z) i \gamma^x \gamma^t \gamma_5 U(z, 0) \psi_q(0)$ with the Wilson line $U(z, 0)= P\exp\left(-ig\int_0^z dz' A_z(z')\right)$ and subscript $q$ as a flavor index.
We calculate the flavor combination $\delta\widetilde{u}-\delta\widetilde{d}$ so that the disconnected diagrams cancel on the lattice.
For the nucleon matrix elements of $\hat{O}(z,a)$ at a given boost momentum, 
$\widetilde{h}(z,P_z,a)$, 
we calculate six source-sink separations $t_\text{sep}\in\{0.54,0.72,0.81,0.90,0.99,1.08\}$~fm, with $\{16,32,32,64,64,64\}$~thousand measurements among 884 gauge configurations, respectively.
Following the work in Ref.~\cite{Bhattacharya:2013ehc}, each three-point correlator, $C_\Gamma^{(3\text{pt})} (P_z, t, t_\text{sep})$ can be decomposed as
\begin{align}
C^\text{3pt}_{\Gamma}(P_z,t,t_\text{sep}) &=
   |{\cal A}_0|^2 \langle 0 | \mathcal{O}_\Gamma | 0 \rangle  e^{-E_0t_\text{sep}} \nonumber\\
   &+|{\cal A}_1|^2 \langle 1 | \mathcal{O}_\Gamma | 1 \rangle  e^{-E_1t_\text{sep}} \nonumber\\
   &+{\cal A}_1{\cal A}_0^* \langle 1 | \mathcal{O}_\Gamma | 0 \rangle  e^{-E_1 (t_\text{sep}-t)} e^{-E_0 t} \nonumber\\
   &+{\cal A}_0{\cal A}_1^* \langle 0 | \mathcal{O}_\Gamma | 1 \rangle  e^{-E_0 (t_\text{sep}-t)} e^{-E_1 t} + \ldots \,,\nonumber
\end{align}
where the source point has been shifted to zero for each measurement, the operator is inserted at time
$t$, and the nucleon state is annihilated at the sink time
$t_\text{sep}$, which (after shifting) is also the source-sink separation.  
The state $|0\rangle$ represents the ground state and $|n\rangle$ with $n
> 0$ the excited states.
In our two-state fits,
the amplitudes ${\cal A}_i$ and the energies $E_i$ are functions of $P_z$ and can be obtained from the corresponding two-point correlators.  
Here we investigate the excited-state contamination by performing fits with and without the 
$\langle 1 | \mathcal{O}_\Gamma | 1 \rangle$  contribution (and labeled them as ``two-simRR'' and ``two-sim'' methods, respectively) and 
using different inputs of source-sink separations $t_\text{sep}$ shown 
in  Fig.~\ref{fig:bareME-tsep}. 
The two-simRR analysis using $t_\text{sep}$ as small as 0.54~fm gives consistent results
with the two-sim analysis using $t_\text{sep}=0.81$~fm, with approximately the same statistical errors after removing the excited-state contamination. Similar results are given by two other fits with larger error as they use fewer three-point proton correlators. Our final result uses the ``two-simRR'' fit with three-point correlators data of $t_\text{sep}$ ranging from 0.72 to 1.08~fm.

\begin{figure}[htbp]
\includegraphics[width=.4\textwidth]{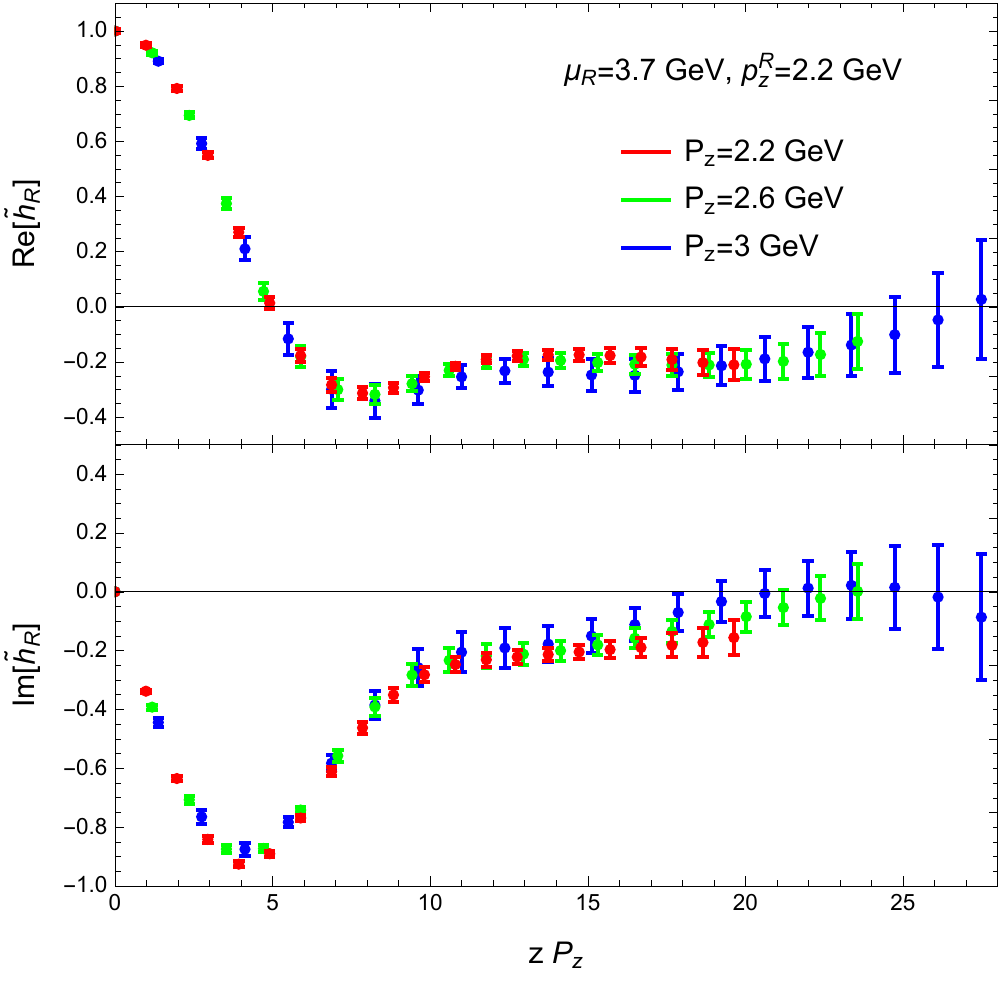}
\caption{The real (top) and imaginary (bottom) parts of the renormalized proton matrix elements as functions of $zP_z$ at renormalization scales {$\mu_R=3.7$~GeV} and $p_z^R=2.2$~GeV. We further normalize all the matrix elements with $h_R(z=0,P_z)$ to improve the signal-to-noise ratios.}
\label{fig:hr}
\end{figure}

%%%%%%%%%%%%%%%%%%
As a second step, we calculate the nonperturbative renormalization (NPR) factor $\widetilde{Z}(z,p^R_z,\mu_R,a)$ from the amputated Green function of $\hat O(z,a)$with a similar procedure to that defined in Ref.~\cite{Liu:2018uuj}, where $p^R_z$ and $\mu_R$ are the Euclidean quark momentum in the $z$-direction and the off-shell quark momentum, respectively.
The bare matrix element of $\hat{O}(z,a)$, $\widetilde{h}(z,P_z,a)$, 
has ultraviolet (UV) power and logarithmic divergences as $a\to 0$ and must be nonperturbatively renormalized to have a well-defined continuum limit. 
%
%The NPR factor $\widetilde{Z}(z,p^R_z,\mu_R,a)$ is calculated using the off-shell quark matrix element of $\hat{O}(z,a)$ where all the loop contributions are subtracted. $\widetilde{Z}$ depends on the Euclidean quark momentum in the $z$-direction, $p_z^R$, and the off-shell momentum squared $\mu_R^2$. 
The NPR factor $\tilde Z(z, p_z^R, \mu_R, a)$ is calculated using the off-shell quark matrix element of $\hat O(z, a)$ and requiring that all the loop corrections are canceled by $\tilde Z(z, p_z^R, \mu_R, a)$ at given $p_z^R$ and $\mu_R$. 
It is computed in Landau gauge using the same lattice ensemble to compute $\widetilde{h}$. 
The renormalized matrix element $\widetilde{h}_R(z,P_z, p^R_z,\mu_R)=\widetilde{Z}^{-1}(z,p^R_z,\mu_R,a)\widetilde{h}(z,P_z,a) $ inherits the dependence on $p_z^R$ and $\mu_R$, which is supposed to be canceled after the later matching step. 
However, since our matching coefficient is only available at one-loop order, there will be remnant %\Blue{two-loop} 
dependence on $p_z^R$ and $\mu_R$ in the final $\delta q(x,\mu)$. On the other hand, the lattice discretization effects of order $O(ap_z^R, a\mu_R)$ or higher are also expected, since we do not take the continuum limit. Both factors will lead to systematic uncertainties in our analysis, so we estimate them by varying the values of $p_z^R$ from 1.3 to 3~GeV, and $\mu_R$ between 2.3 and 3.7~GeV. Our results show an insensitivity to $\mu_R$ but noticeable dependence on $p_z^R$. We choose $\widetilde{h}_R(z,P_z, p^R_z,\mu_R)$ at $\mu_R=3.7$~GeV and $p_z^R=2.2$~GeV to be the central value, as shown in Fig.~\ref{fig:hr}. We include the variation of $\mu_R$ and $p_z^R$ as sources of systematic uncertainties.

%%%%%%%%%%%%%%%%%%

Next, we Fourier transform the $\widetilde{h}_R(z,P_z, p^R_z,\mu_R)$ into $x$-space to obtain the quasi-distribution $\delta\widetilde{q}(x, P_z, p_z^R, \mu_R)$. 
As shown in Fig.~\ref{fig:hr}, the long-range correlation which dominates the small-$x$ distribution has much larger statistical uncertainty, and the higher-twist effects as well as finite-volume effects will also become important with larger $|z|$. 
Therefore, we have to truncate the Fourier transform at a finite $|z_\text{max}|$, which will limit our prediction for the small-$x$ distribution and introduce an unphysical oscillation in $x$-space 
%\sout{as a numerical effect}
which can be removed by using the ``derivative'' method proposed in our earlier work~\cite{Lin:2017ani}:
$\delta \tilde{q}(x,P_z,p_z^R,\mu_R) = i\int_{-z_\text{max}}^{+z_\text{max}} \!\!\! dz e^{i x P_z z} \tilde{h}'_R(z,P_z,p_z^R,\mu_R)/x$. 
In this work, we vary $|z_\text{max}|$ to estimate the remaining corresponding error, which turns out to be small compared with other systematics. 
In our final result, we choose $z_\text{max}P_z\approx 20$. 

%%%%%%%%%%%%%%%%%%%%%%%%%
{\em From quasi-PDF to physical PDF:}
The transversity quasi-PDF $\delta\widetilde{q}(x, P_z, p_z^R, \mu_R)$ in the RI/MOM scheme is related to the $\overline{\text{MS}}$ physical lightcone transversity PDF $\delta q(x,\mu)$ at scale $\mu$ through the following factorization formula~\cite{Stewart:2017tvs,Liu:2018uuj}:
\begin{align} \label{eq:matching}
\delta\widetilde{q}(x, P_z, p_z^R, \mu_R) =& \int_{-1}^1{dy \over |y|} C\left({x\over y},{\mu_R\over p_z^R},{yP_z \over \mu},{yP_z\over p_z^R}\right) \delta q(y,\mu)\nn\\
&+ O\left({M^2\over P_z^2},{\Lambda^2_\text{QCD}\over P_z^2}\right)\,,
\end{align}
where $M$ is the proton mass, $C$ is the perturbative matching coefficient, and the antiquark distribution $\delta \bar{q}(y,\mu)=-\delta q(-y,\mu)$ is rearranged into the region $-1<y<0$.

Since the tensor charge for the transversity PDF is not conserved, the matching coefficient $C$ in Eq.~\ref{eq:matching} cannot be written as a plus function to enforce charge conservation as in the unpolarized and helicity cases~\cite{Stewart:2017tvs,Liu:2018uuj}.
Instead, the lowest moment of $C$ is not unity and must be consistent with the anomalous dimension of the tensor charge. The details of treating this subtlety as well as the complete one-loop result for $C$ in Landau gauge is available in the Appendix.
For comparison, we note that ETMC used a two-step matching procedure where they first converted the RI/MOM transversity quasi-PDF into the $\overline{\text{MS}}$ scheme, and then matched the latter onto $\delta q(x,\mu)$~\cite{Alexandrou:2018eet}.
%However, their matching coefficient for the $\overline{\text{MS}}$ transversity quasi-PDF is given as a plus function, indicating that they may have not consistently treated the subtlety about charge non-conservation.

\begin{figure}[htbp]
\includegraphics[width=.4\textwidth]{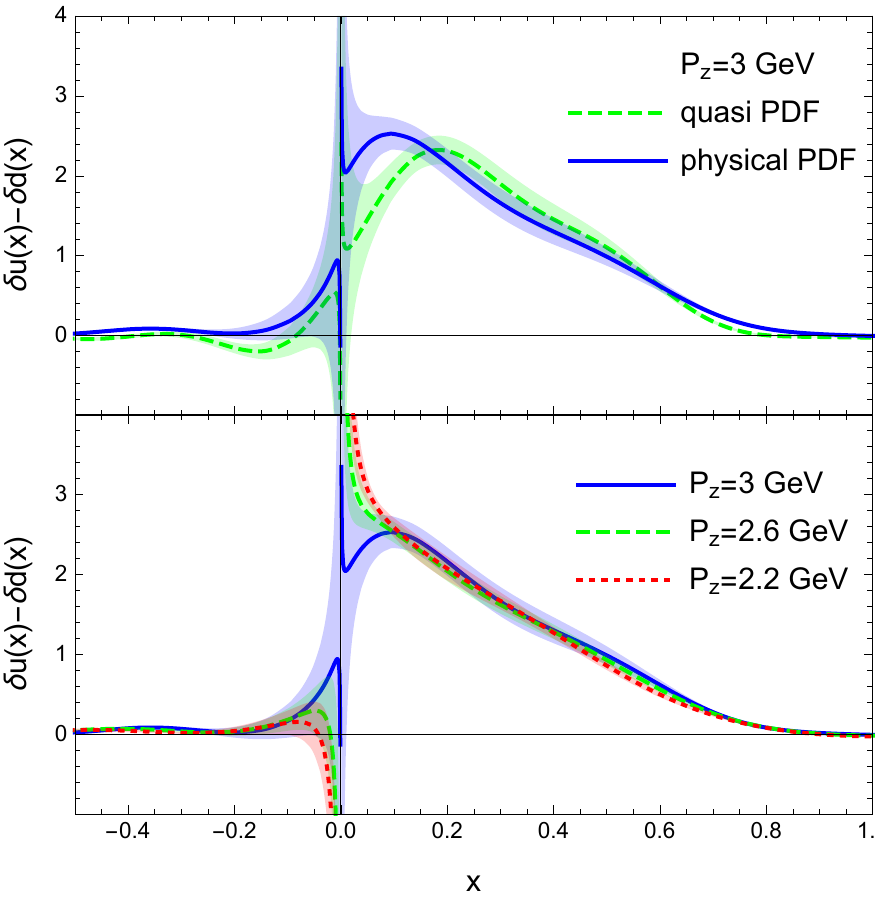}
\caption{(Top) The quark transversity quasi-PDF in RI/MOM scheme at proton momentum 3.0~GeV and resulting physical PDF in $\overline{\text{MS}}$ scheme at  $\sqrt{2}$~GeV.
(Bottom) The matched physical PDFs for all three proton momenta. The error bands are statistical only.} \label{fig:matchingPDFallPz}
\end{figure}

We obtain the final result for the lightcone transversity $\delta q(x,\mu)$ after applying matching to the quasi-PDF $\delta \tilde{q}(x,P_z,p_z^R,\mu_R)$ and the mass correction~\cite{Chen:2016utp}.
A comparison between the RI/MOM-renormalized quasi-PDF calculated at $P_z=3.0$~GeV and the physical PDF is shown in the upper panel of Fig.~\ref{fig:matchingPDFallPz}, where only the statistical errors are plotted. 
The physical transversity is normalized to $g_T = 0.99(4)$, calculated in Ref.~\cite{Gupta:2018qil}. 
While the mass correction is negligible, the matching suppresses the quasi-distribution in mid-$|x|$ range $0.2<x<0.6$, and enhances it in the small-$|x|$ region $-0.1<x<0.2$.
The antiquark distribution is small in both the quasi- and matched PDFs. 
Note that the current distribution still contains higher-twist contributions of $O(\Lambda_\text{QCD}^2/P_z^2)$, whose effect can be assessed by varying $P_z$ in the lattice calculation.
In the lower panel of Fig.~\ref{fig:matchingPDFallPz}, we plot the physical PDFs $\delta q(x,\mu)$ at $\mu=\sqrt{2}$~GeV from calculations at all three proton momenta. As one can see, for $|x|>0.1$, the results converge as we increase the proton momentum $P_z$, which indicates that the higher-twist effects are well suppressed by the large momentum. Moreover, the results for $x<-0.2$ are all consistent with zero, which offers a strong constraint on the anti-sea flavor asymmetry. As for $|x|<0.1$, the final result is susceptible to the nucleon momentum, but it is also sensitive to the systematic uncertainties in the long-range correlations which are not well controlled.

\begin{figure}[htbp]
\includegraphics[width=.5\textwidth]{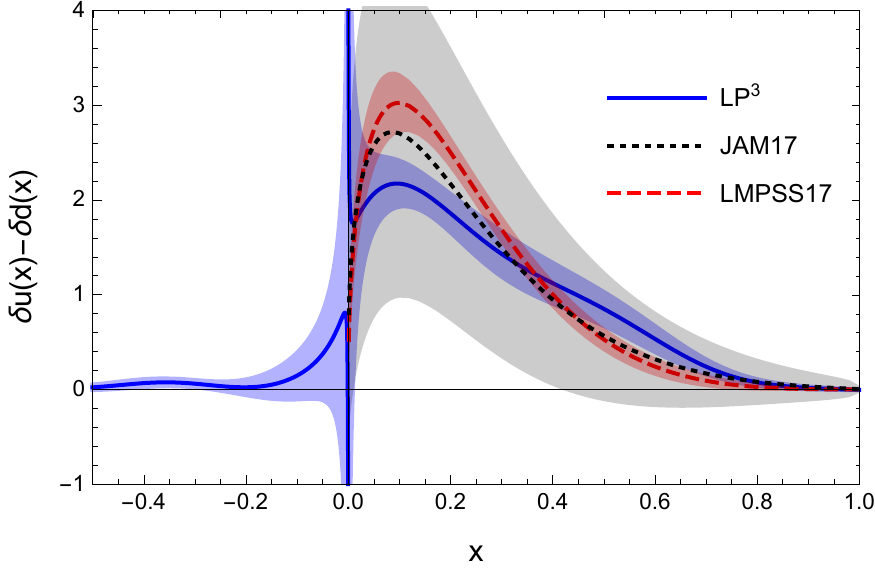}
\caption{Our final proton isovector transversity PDF at renormalization scale $\mu=\sqrt{2}$~GeV (
$\overline{\text{MS}}$ scheme), extracted from lattice QCD and LaMET at $P_z=3$~GeV, compared with global fits by JAM17 and LMPSS17~\cite{Lin:2017stx}.
The blue error band includes statistical errors (which fold in the excited-state uncertainty) and systematics.}\label{fig:finalPDF}
\end{figure}

Our final result (taken from the physical transversity with $P_z=3.0$~GeV) is compared with JAM17 and LMPSS17~\cite{Lin:2017stx} in Fig.~\ref{fig:finalPDF}. 
The systematics of Fig.~\ref{fig:finalPDF} are of two types: 
First, using the continuum and infinite-volume extrapolations of $g_T$ in Ref.~\cite{Gupta:2018qil}, 
which includes the same lattices used in this work, 
we conservatively estimate our lattice systematics due to $g_T$ used in normalization ($4\%$), 
finite volume ($0.5\%$), 
nonzero lattice spacing (enlarged from $a$ to $3a$ to compensate the length contraction of a moving proton) ($3.5\%$), 
which add up to $5.4\%$. 
Second, the renormalization scale dependence in the NPR procedure, 
which is the dominant systematic. 
These two types of systematics are combined in quadrature to give our final result. 
Our prediction for $\delta q(x,\mu)$ is consistent with the global analysis of JAM17 but with a significantly smaller uncertainty and an unequivocal positive-definite isovector distribution over the range $0<x<1$. Compared to the  LMPSS17 constrained fit using lattice averaged $g_T$, our result shows a remarkable agreement within $2\sigma$. 
Moreover, in contrast to the recent calculation by ETMC~\cite{Alexandrou:2018eet}, our antiquark distribution favors a vanishing
flavor asymmetry in the antiquark sea, which has been assumed in all global analyses so far. 
Based on our earlier works~\cite{Chen:2017mzz,Lin:2017ani,Chen:2017lnm}, we speculate that this is due to a systematic error from the truncation in the Fourier transformation. 
As shown in Fig.~4 of Ref.~\cite{Lin:2017ani}, the antiquark region is highly sensitive to the proton momentum used in the calculation. In this work, our proton momentum $P_z=3.0$~GeV is about factor of 2 larger than that in ETMC's (1.4~GeV), which could also explain the different results.

To summarize, we have calculated the isovector quark transversity PDF with LaMET at physical pion mass and large nucleon momenta. With high statistics, we have performed multi-state analyses with multiple source-sink separations to remove the excited-state contamination, and added systematic corrections to our lattice matrix elements to obtain the final result. We have reached a precision that is significantly better than the up-to-date global analysis, and provided the first clear evidence to support the flavor symmetry in anti-sea distribution. For future calculations, we will use smaller lattice spacings to take the continuum limit and reach higher proton momenta, so that we can further reduce our systematics and offer more insights on the small-$x$ distributions. The present calculation is also an exemplary case where a LaMET calculation of the PDFs can advance current experiments. Our result will provide guidance to the relevant experiments at JLab 12-GeV and FNAL at present, as well as EIC in the future.

\section*{Acknowledgments}
We thank the MILC Collaboration for sharing the lattices used to perform this study. The LQCD calculations were performed using the multigrid algorithm~\cite{Babich:2010qb,Osborn:2010mb} and Chroma software
suite~\cite{Edwards:2004sx}.
This research used resources of the National Energy Research Scientific Computing Center, a DOE Office of Science User Facility supported by the Office of Science of the U.S.
Department of Energy under Contract No. DE-AC02-05CH11231
through ALCC and ERCAP;
facilities of the USQCD Collaboration, which are funded by the Office of Science of the U.S. Department of Energy,
and supported in part by Michigan State University through computational resources provided by the Institute for Cyber-Enabled Research.
HL, RL, and YY are supported by the US National Science Foundation under grant PHY 1653405 ``CAREER: Constraining Parton Distribution Functions for New-Physics Searches''.
JWC is partly supported by the Ministry of Science and Technology, Taiwan, under Grant No. 105-2112-M-002-017-MY3 and the Kenda Foundation. LJ is supported by the Department
of Energy, Laboratory Directed Research and Development (LDRD) funding of BNL, under contract DE-EC0012704. 
YSL is supported by Science and Technology Commission of Shanghai Municipality (Grant No.16DZ2260200) and National Natural Science Foundation of China (Grant No.11655002).  JZ is supported by the SFB/TRR-55 grant ``Hadron Physics from
Lattice QCD'', and a grant from National Science Foundation of China (No.~11405104). YZ is supported by the U.S. Department of Energy, Office of Science, Office of Nuclear
Physics, from DE-SC0011090 and within the framework of the TMD Topical Collaboration.

%\bibliographystyle{apsrev4-1}
%\bibliography{ref}


\begin{widetext}

\clearpage
\appendix

\section{Matching the transversity quasi-PDF in the RI/MOM scheme}	

The transversity quasi-PDF is defined as
\begin{equation}
\delta\tilde{q}_\Gamma(x, P_z, \tilde{\mu})
= \int_{-\infty}^\infty \frac{dz}{2\pi}\ e^{ixP_zz} \big\langle P \big| \bar{\psi}(z)\Gamma U(z,0)\psi(0)\big|P\big\rangle\,,
\end{equation}
where $U(z,0)= P\exp\left(-ig\int_0^z dz' A_z(z')\right)$ is the spacelike Wilson line and $\Gamma=\frac{1}{2}\gamma^x\gamma^t\gamma_5$.

\subsection{One-loop corrections to the lightcone and quasi-PDFs}

In general covariant gauge, the gluon propagator is
\begin{equation}
iD_\tau^{\mu\nu}(k)=-\frac{i}{k^2}\left[g^{\mu\nu}-(1-\tau)\frac{k^\mu k^\nu}{k^2}\right].
\end{equation}
The Landau gauge corresponds to $\tau=0$. The matrix element of the transversity lightcone and quasi-PDFs are calculated in an off-shell quark state with momentum $p^\mu$ ($p^2<0$).

\subsubsection{Lightcone PDF in the $\overline{\text{MS}}$ scheme}

At tree-level, the matrix element of the lightcone PDF is
\begin{equation}
\delta q^{(0)}(x)=\delta(1-x)\,.
\end{equation}
%\begin{align}
%q^{(0)}=\frac{\bar{u}(p)\gamma^x\gamma^+\gamma_5u(p)}{2p^+},
%\end{align}
In the $\overline{\text{MS}}$ scheme, the one-loop correction is
\begin{align}
\delta q^{(1)} (x,\mu,p) &= \delta q^{(1)}_\text{real}(x,\mu,p) + \delta q^{(1)}_\text{virtual}(x,\mu,p)\,,\\
\delta q^{(1)}_\text{real}(x,\mu,p)&=\frac{\alpha_s C_F}{2\pi}\left[\frac{2x}{1-x}\ln\frac{\mu^2}{-x(1-x)p^2}-(1-\tau)\frac{x}{2(1-x)}\right]\,,\label{eq:LCPDF_r}\\
\delta q^{(1)}_\text{virtual}(x,\mu,p)&=\delta(1-x)\left\{\delta q^{(1)}_\text{wf}(-p^2)-\frac{\alpha_s C_F}{2\pi}\int_{-1}^1 dx'\left[\frac{2x'}{1-x'}\ln\frac{\mu^2}{-x'(1-x')p^2}-(1-\tau)\frac{x'}{1-x'}\right]\right\}\,.\label{eq:LCPDF_v}
\end{align}
where the wavefunction renormalization contributes
\begin{align}\label{eq:LCPDF_wf}
\delta q^{(1)}_\text{wf}(-p^2)=-\tau\frac{\alpha_sC_F}{4\pi}\left(\ln\frac{\mu^2}{-p^2}+1\right)\,.
\end{align}
Note that $\delta q^{(1)}_{wf}=0$ in the Landau gauge.

\subsubsection{Quasi-PDF in the RI/MOM scheme}

At tree level, the matrix element of the quasi-PDF is
\begin{equation}
\delta\widetilde{q}^{(0)}(x)=\delta(1-x)\,.
\end{equation}

In dimensional regularization ($d=4-2\epsilon$), the one-loop correction to the bare quasi-PDF is
\begin{equation}\label{eq:quasibare}
\tilde{q}^{(1)}(x,p,\epsilon) = \tilde{q}^{(1)}_\text{real}(x,p) + \tilde{q}^{(1)}_\text{virtual}(x,p,\epsilon)
\end{equation}

\begin{align}
\delta\tilde{q}^{(1)}_\text{real}(x,p)
=&\frac{\alpha_s C_F}{2\pi}\Bigg\{\left[-\frac{1}{|1-x|}+\frac{x}{1-x}F_{1/2}^1+(1-\tau)\frac{\rho}{8}\left(F_{3/2}^{y-y^2}-3F_{5/2}^{y(1-y)[x^2-y^2(1-\rho)]}\right)\right]\gamma^x\gamma^t\nonumber\\
&+\left[\frac{1}{1-x}F_{1/2}^y+(1-\tau)\frac{3\rho}{4}F_{5/2}^{y^2(1-y)(x-y)}\right]\left(\frac{p_t}{p_z}\gamma^x\gamma^z-\frac{p_x}{p_z}\gamma^t\gamma^z\right)\nonumber\\
&-(1-\tau)\frac{1}{4}\left[F_{3/2}^{y-y^2}-3F_{5/2}^{y(1-y)(x-y)^2}\right]\left(\frac{p_t}{p_z^2}\slashed{p}\gamma^x-\frac{p_x}{p_z^2}\slashed{p}\gamma^t\right)\Bigg\}\gamma_5 \,,
\end{align}
\begin{align}
\delta\tilde{q}^{(1)}_\text{virtual}(x,p,\epsilon)
=&\delta(1-x)\frac{\alpha_s C_F}{2\pi}\Bigg\{\left[-{\tau\over2\epsilon'}+\delta q^{(1)}_\text{wf}(-p^2)\right]\gamma^x\gamma^t\nonumber\\
&-\int_{-\infty}^\infty dx'\left[-\frac{1}{|1-x'|}+\frac{x'}{1-x'}F_{1/2}^1+(1-\tau)\frac{1}{2}\left(\frac{1}{|1-x'|}-F_{3/2}^{y[x'-y(1-\rho)]}\right)\right]\gamma^x\gamma^t\nonumber\\
&-\int_{-\infty}^\infty dx'\left[\frac{1}{1-x'}F_{1/2}^y+(1-\tau)\frac{1}{2}F_{3/2}^{y(x'-y)}\right]\left(\frac{p_t}{p_z}\gamma^x\gamma^z-\frac{p_x}{p_z}\gamma^t\gamma^z\right)\Bigg\}\gamma_5 \,,
\end{align}
where
\begin{align}
{1\over \epsilon'} ={1\over\epsilon} -\gamma_E +\ln(4\pi),\qquad \rho=\frac{-p^2-i\varepsilon}{p_z^2}\,,
\end{align}
where $i\varepsilon$ gives the prescription to analytically continue $\rho$ from $\rho<1$ to $\rho>1$. The $F$'s are defined in the appendix. $\bar{u}(p)\cdots u(p)$ is defined by the projection for the off-shell matrix element~\cite{Liu:2018uuj}.

According to Ref.~\cite{Stewart:2017tvs}, to match to the lightcone PDF, one has to take the on-shell ($-p^2\to0$) and collinear ($p_t\to p_z$ and $p_x\to 0$) limits of the bare quasi-PDF matrix element. We observe that both the $\gamma^x\gamma^t\gamma_5$ and $\gamma^x\gamma^z\gamma_5$ terms approach the lightcone operator ($\Gamma=\gamma^x\gamma^+\gamma_5$) in this limit. So the combination of these two terms in Eq.~\ref{eq:quasibare} gives the correct collinear divergence $\ln(-p^2)$. Therefore, we sum the coefficients of $\gamma^x\gamma^t\gamma_5$ and $\gamma^x\gamma^z\gamma_5$ to obtain the bare quasi-PDF 
\begin{align}
\delta\tilde{q}^{(1)}_{B,\text{real}}(x,\rho)=&\frac{\alpha_s C_F}{2\pi} \left\{
\begin{array}{lc}
\displaystyle\frac{2x}{1-x}\ln\frac{x}{x-1} & x>1\\
\displaystyle-\frac{2x}{1-x}-(1-\tau)\frac{x}{2(1-x)}+\frac{2x}{1-x}\ln\frac{4}{\rho} & 0<x<1\\
\displaystyle-\frac{2x}{1-x}\ln\frac{x}{x-1} & x<0
\end{array}
\right.\,,\label{eq:bare_PDF_r}\\
\delta\tilde{q}^{(1)}_{B,\text{virtual}}(x,\rho,\epsilon)=&\delta(1-x)\frac{\alpha_s C_F}{2\pi}\left[-{\tau\over2\epsilon'}+\delta q^{(1)}_\text{wf}(-p^2)\right] \nonumber\\
&-\delta(1-x)\frac{\alpha_s C_F}{2\pi} \int_{-\infty}^\infty dx'\left\{
\begin{array}{lc}
\displaystyle-(1-\tau)\frac{1}{2(1-x')}+\frac{2x'}{1-x'}\ln\frac{x'}{x'-1} & x'>1\\
\displaystyle-\frac{2x'}{1-x'}+(1-\tau)\frac{1-2x'}{2(1-x')}+\frac{2x'}{1-x'}\ln\frac{4}{\rho} & 0<x'<1\\
\displaystyle(1-\tau)\frac{1}{2(1-x')}-\frac{2x'}{1-x'}\ln\frac{x'}{x'-1} & x'<0
\end{array}
\right.\,.\label{eq:bare_PDF_v}
\end{align}

The RI/MOM renormalization condition is
\begin{align}
Z(z,p^R_z,a^{-1},\mu_R)=\left.\frac{\sum_s\langle p,s|O_{\Gamma}(z)|p,s\rangle}{\sum_s\langle p,s|O_{\Gamma}(z)|p,s\rangle_\text{tree}}\right|_{\tiny\begin{matrix}p^2=-\mu_R^2 \\ \!\!\!\!p_z=p^R_z\end{matrix}}
\end{align}
Only the terms proportional to $\gamma^x\gamma^t\gamma_5$ contribute to the ultraviolet (UV) divergence, so we choose the minimal projection~\cite{Liu:2018uuj} to define the renormalization counterterm as the coefficient of $\gamma^x\gamma^t\gamma_5$,
%\begin{align}
%\tilde{q}^{(1)}_{CT}(x,r)=\tilde{q}^{(1)}(x,p_z=p_R^z,p^2=-\mu_R^2)
%\end{align}
\begin{align}
&\delta\tilde{q}^{(1)}_{\text{CT},\text{real}}(x,p_z^R,\mu_R)=\frac{\alpha_s C_F}{2\pi} \label{eq:CT_PDF_r}\\
&\times\left\{
\begin{array}{lc}
\displaystyle\frac{1}{1-x}+(1-\tau)\frac{r(1+r-6x+4x^2)}{2(r-1)(1-x)(r-4x+4x^2)}-\left[\frac{2x}{1-x}+(1-\tau)\frac{r}{r-1}\right]\frac{1}{\sqrt{r-1}}\tan^{-1}\frac{\sqrt{r-1}}{1-2x} & x>1\\
\displaystyle-\frac{1}{1-x}+(1-\tau)\frac{-1-r+2x}{2(r-1)(1-x)}+\left[\frac{2x}{1-x}+(1-\tau)\frac{r}{r-1}\right]\frac{1}{\sqrt{r-1}}\tan^{-1}\sqrt{r-1} & 0<x<1\\
\displaystyle-\frac{1}{1-x}-(1-\tau)\frac{r(1+r-6x+4x^2)}{2(r-1)(1-x)(r-4x+4x^2)}+\left[\frac{2x}{1-x}+(1-\tau)\frac{r}{r-1}\right]\frac{1}{\sqrt{r-1}}\tan^{-1}\frac{\sqrt{r-1}}{1-2x} & x<0
\end{array}
\right.\,,\nonumber\\
&\delta\tilde{q}^{(1)}_{\text{CT},\text{virtual}}(x,p_z^R,\mu_R,\epsilon)=\delta(1-x)\frac{\alpha_s C_F}{2\pi}\left[-{\tau\over2\epsilon'}+\delta q^{(1)}_\text{wf}(\mu_R^2)\right]-\delta(1-x)\frac{\alpha_s C_F}{2\pi}\int_{-\infty}^\infty dx'\label{eq:CT_PDF_v}\\
&\times\left\{
\begin{array}{lc}
\displaystyle\frac{1}{1-x'}+(1-\tau)\frac{r}{2(1-x')(r-4x'+4x'^2)}-\left[\frac{2x'}{1-x'}+(1-\tau)\right]\frac{1}{\sqrt{r-1}}\tan^{-1}\frac{\sqrt{r-1}}{1-2x'} & x'>1\\
\displaystyle-\frac{1}{1-x'}-(1-\tau)\frac{1}{2(1-x')}+\left[\frac{2x'}{1-x'}+(1-\tau)\right]\frac{1}{\sqrt{r-1}}\tan^{-1}\sqrt{r-1} & 0<x'<1.\\
\displaystyle-\frac{1}{1-x'}-(1-\tau)\frac{r}{2(1-x')(r-4x'+4x'^2)}+\left[\frac{2x'}{1-x'}+(1-\tau)\right]\frac{1}{\sqrt{r-1}}\tan^{-1}\frac{\sqrt{r-1}}{1-2x'} & x'<0
\end{array}
\right.\,,\nonumber
\end{align}
where $r\equiv{\mu_R}^2/(p_z^R)^2$, and $\mu_R^2\ge (p_z^R)^2$.

The renormalized one-loop quasi-PDF in RI/MOM scheme is
\begin{align}
\delta\tilde{q}_{R,\text{real}}^{(1)}\left(x,\rho,p_z^R,\mu_R\right)&=\delta\tilde{q}^{(1)}_{B,\text{real}}(x,\rho)-\frac{|p_z|}{p_z^R}\delta\tilde{q}^{(1)}_{\text{CT},\text{real}}\left(1+\frac{p_z}{p_z^R}(x-1),p_z^R,\mu_R\right)\label{eq:renorm_PDF_r}\\
\delta\tilde{q}_{R,\text{virtual}}^{(1)}(x,\rho,p_z^R,\mu_R)&=\delta\tilde{q}^{(1)}_{B,\text{virtual}}(x,\rho,\epsilon)-\delta\tilde{q}^{(1)}_{\text{CT},\text{virtual}}(x,p_z^R,\mu_R,\epsilon).\label{eq:renorm_PDF_v}
\end{align}
Note that the renormalized wavefunction contribution is
\begin{align}\label{eq:renorm_PDF_wf}
\delta q^{(1)}_\text{wf}(-p^2)-\delta q^{(1)}_\text{wf}(\mu_R^2)=\frac{\alpha_s C_F}{2\pi}\frac{\tau}{2}\ln\frac{r}{\rho}\,,
\end{align}
which vanishes in the Landau gauge.

\subsection{Matching Coefficient}
The matching between quasi-PDF $\tilde{q}(x,P_z, p^R_z,\mu_R)$ and the $\overline{\text{MS}}$ PDF $q(y,\mu)$ at scale $\mu$ is
\begin{align}
\delta\tilde{q}(x,P_z,p^R_z,\mu_R)=\int_{-1}^1\frac{dy}{|y|}C\left(\frac{x}{y},r,\frac{yP_z}{\mu},\frac{yP_z}{p_z^R}\right) \, \delta q(y,\mu)+\mathcal{O}\left({M^2\over P_z^2},{\Lambda_{\text{QCD}}^2\over P_z^2}\right),
\end{align}
where $r={\mu_R}^2/{p^R_z}^2$, and we have absorbed the antiquark distribution $\bar{q}(y,\mu)=-q(-y,\mu)$ into the region $-1<y<0$. 
The matching coefficient $C$ at one-loop level is
\begin{align}
C\left(x,r,\frac{p_z}{\mu},\frac{p_z}{p_z^R}\right)=\delta(1-x)+\left[\delta\tilde{q}_{R,\text{real}}^{(1)}+\delta\tilde{q}_{R,\text{virtual}}^{(1)}-\delta q_\text{real}^{(1)}-\delta q_\text{virtual}^{(1)}\right]+\mathcal{O}\left(\alpha_s^2\right)\,.
\end{align}

%\appendix
\subsection{Definitions of $F$'s}

The general definition of the $F^X_n$ is given by
\begin{align}
F^X_n=\int_0^1 dy \frac{X}{[(x-y)^2+y(1-y)\rho]^n}\,.
\end{align}

If $\rho<1$,
\begin{align}
F^{1}_{1/2}&=\frac{1}{\sqrt{1-\rho}}\ln\frac{|1-x|+|x|+\sqrt{1-\rho}}{|1-x|+|x|-\sqrt{1-\rho}},\,\,
F^{y}_{1/2}=\frac{|1-x|-|x|}{1-\rho}-\frac{\rho-2x}{2(1-\rho)}F^1_{1/2}\,,\\
F^{1}_{3/2}&=\frac{2}{\rho}\frac{(\rho-2+2x)|x|+(\rho-2x)|1-x|}{(\rho-4x+4x^2)|x(1-x)|},\,\,
F^{y}_{3/2}=\frac{2}{\rho}\frac{\rho-2x(1-x)-2|x(1-x)|}{(\rho-4x+4x^2)|1-x|}\,,\\
F^{y^2}_{3/2}&=-\frac{2}{\rho}\frac{\rho^2-\rho x(4-3x)+2x^2(1-x)+(2x-\rho)|x(1-x)|}{(1-\rho)(\rho-4x+4x^2)|1-x|}+\frac{1}{1-\rho}F^1_{1/2}\,,\\
F^{y(1-y)}_{5/2}&=\frac{4}{3\rho^2} \left[\frac{\rho^2-4\rho x+8x^2-8x^3}{(\rho-4x+4x^2)^2|x|}+\frac{\rho^2-4\rho+4\rho x+8x(1-x)^2}{(\rho-4x+4x^2)^2|1-x|}\right]\,,\\
F^{y^2(1-y)}_{5/2}&=\frac{4}{3\rho^2}\left[\frac{-4x^2(\rho-2x+2x^2)}{(\rho-4x+4x^2)^2|x|}+\frac{\rho^2-4\rho x+4\rho x^2+8x^2(1-x)^2}{(\rho-4x+4x^2)^2|1-x|}\right]\,.
\end{align}
If $\rho>1$, we need to take analytic continuation
\begin{align}
F^{1}_{1/2}&=\frac{2}{\sqrt{\rho-1}}\tan^{-1}\frac{\sqrt{\rho-1}}{|1-x|+|x|}.
\end{align}

\subsection{Loop Integrals}
The integration region for all loop integrals are
\begin{equation}
\int\frac{d^4k}{(2\pi)^4}2\pi\delta(k^z-xp^z)\,.
\end{equation}
Then the loop integrals can be calculated as the $F$'s defined above, and shown in the following table:
\vspace{.5cm}

\centering
\scalebox{1.2}{
\begin{tabular}{|c|c||c|c||c|c|}
\hline
Integrand & Result & Integrand & Result & Integrand & Result\\
\hline
$\frac{1}{k^2}$ & $\frac{i}{4\pi}|x|p^z$ & $\frac{1}{k^2(k-p)^2}$ & $\frac{i}{8\pi}\frac{1}{p^z}F^1_{1/2}$ & $\frac{k^t}{k^2(k-p)^2}$ &$\frac{i}{8\pi}\frac{p^t}{p^z}F^y_{1/2}$\\
\hline
$\frac{1}{k^4}$ & $\frac{i}{8\pi}\frac{1}{|x|p^z}$ & $\frac{1}{k^2(k-p)^4}$ & $-\frac{i}{16\pi}\frac{1}{{p^z}^3}F^{y}_{3/2}$ & $\frac{k^t}{k^2(k-p)^4}$ & $-\frac{i}{16\pi}\frac{p^t}{{p^z}^3}F^{y^2}_{3/2}$ \\
\hline
$\frac{1}{(k-p)^2}$ & $\frac{i}{4\pi}|1-x|p^z$ & $\frac{1}{k^4(k-p)^2}$ & $-\frac{i}{16\pi}\frac{1}{{p^z}^3}F^{1-y}_{3/2}$ & $\frac{k^t}{k^4(k-p)^2}$ & $-\frac{i}{16\pi}\frac{p^t}{{p^z}^3}F^{y(1-y)}_{3/2}$ \\
\hline
$\frac{1}{(k-p)^4}$ & $\frac{i}{8\pi}\frac{1}{|1-x|p^z}$ & $\frac{1}{k^4(k-p)^4}$ & $\frac{3i}{32\pi}\frac{1}{{p^z}^5}F^{y(1-y)}_{5/2}$ & $\frac{k^t}{k^4(k-p)^4}$ & $\frac{3i}{32\pi}\frac{p^t}{{p^z}^5}F^{y^2(1-y)}_{5/2}$ \\
\hline
\end{tabular}
}
\vspace{1cm}

\end{widetext} 

\end{document}